\documentclass[conference,compsoc]{IEEEtran}
\IEEEoverridecommandlockouts
\IEEEpubid{\begin{minipage}{\textwidth}\centering\footnotesize
This work has been submitted to the IEEE for possible publication. Copyright may be
transferred without notice, after which this version may no longer be accessible.
\end{minipage}}

\ifCLASSOPTIONcompsoc
  \usepackage[nocompress]{cite}
\else
  \usepackage{cite}
\fi
\usepackage[T1]{fontenc}
\usepackage[utf8]{inputenc}
\usepackage{amsmath,amssymb,amsfonts}
\usepackage{graphicx}
\usepackage{booktabs}
\usepackage{textcomp}
\usepackage{xcolor}
\usepackage{url}
\usepackage[hidelinks,pdfauthor={},pdfcreator={},pdfproducer={}]{hyperref}

\newif\ifanonymous
\anonymousfalse

\begin{document}

\title{Intent-Aware Protection for AI Agents}

\ifanonymous
\author{\IEEEauthorblockN{Anonymous Submission}}
\else
\author{
\IEEEauthorblockN{Fady Copty, Neta Haiby, Max Braitmaiere, Julia Korsunsky}
\IEEEauthorblockA{Microsoft}
}
\fi

\maketitle

\begin{abstract}
AI agents increasingly access enterprise data, invoke powerful tools, and act with limited human oversight. Existing security defenses primarily detect prompt injection, harmful outputs, or suspicious behavior, but do not determine whether an agent remains faithful to the purpose for which it was deployed. An action may contain no attack pattern and stay within granted permissions while still violating the agent's role, organizational constraints, or purpose. We introduce Intent-Aware Security, a framework that treats trusted agent intent as a first-class security object. We define Intent Preservation and Intent Drift, and operationalize them through an Intent Profile: an agent-specific representation extracted from organizational, role, and developer intent. We further introduce an Intent-Aware benchmark of 887 validated samples across 35 agents, built by a generation procedure that fixes ground-truth labels before LLM realization, and four Intent Profile-based protections: pre-execution detection, post-execution detection, input alignment, and system-prompt hardening.

Across all evaluations, Intent Profile-based protection outperforms all profile-free baselines on the primary metric. On our benchmark, it raises four-class macro-F1 from 45\% to 79\% (Out-of-Scope F1 from 8\% to 76\%). On more than 10,000 executed AgentDojo trajectories, where ground truth comes from environment state, profile-based post-execution detection reaches 0.949 AUROC versus AlignmentCheck~0.909, 0.897 with raw agent configuration and 0.843 without context. Profile-based input alignment improves protection success by 7 points over DataFilter with a comparable benign utility. These results suggest that an agent's intended purpose can serve as a useful security reference for detecting and mitigating deviations from its deployed role and operational boundaries.
\end{abstract}

\section{Introduction}\label{sec:introduction}

Large Language Model (LLM)-based agents are increasingly moving beyond conversational assistance toward systems that access enterprise data, invoke tools, plan multi-step tasks, and take actions on behalf of users. As agents gain greater capability and autonomy, securing their behavior becomes increasingly important. Existing defenses address important threats such as prompt injection, harmful outputs, data leakage, and unauthorized tool use. However, these defenses do not fully address a different question: \emph{is the requested or resulting behavior appropriate for the specific purpose for which the agent was deployed?}

The distinction matters because the security interpretation of an interaction can depend on the agent receiving it. A request may contain no recognizable attack pattern and require only authorized capabilities yet still be inconsistent with the agent's intended purpose. Conversely, content that appears adversarial in isolation may be legitimate for an agent whose purpose is to analyze security threats. For example, asking an AI security agent to analyze a prompt-injection payload may be legitimate, whereas directing a travel-planning agent to follow that same payload would not be. Security decisions based only on the interaction therefore lack information about whether the requested behavior is appropriate for the particular agent.

This observation motivates a central research question of this paper:

\emph{Does conditioning security decisions on an explicit representation of an agent's intent improve our ability to distinguish legitimate use, scope violations, agent repurposing, and attacks?}

We investigate this question through \textbf{Intent-Aware Security}, an approach that makes the intended purpose and operational boundaries of a deployed agent explicit inputs to security decisions. We define \textbf{Intent Preservation} as the property that an agent's behavior remains consistent with the organizational, role, and developer constraints governing its deployment. We distinguish this property from \textbf{Intent Drift}, which occurs when the agent's resulting behavior violates those constraints. Importantly, a request that attempts to induce a violation is not itself Intent Drift: drift occurs only if the agent's resulting behavior departs from its intended boundaries.

\begin{figure*}[t]
\centering
\includegraphics[width=\textwidth]{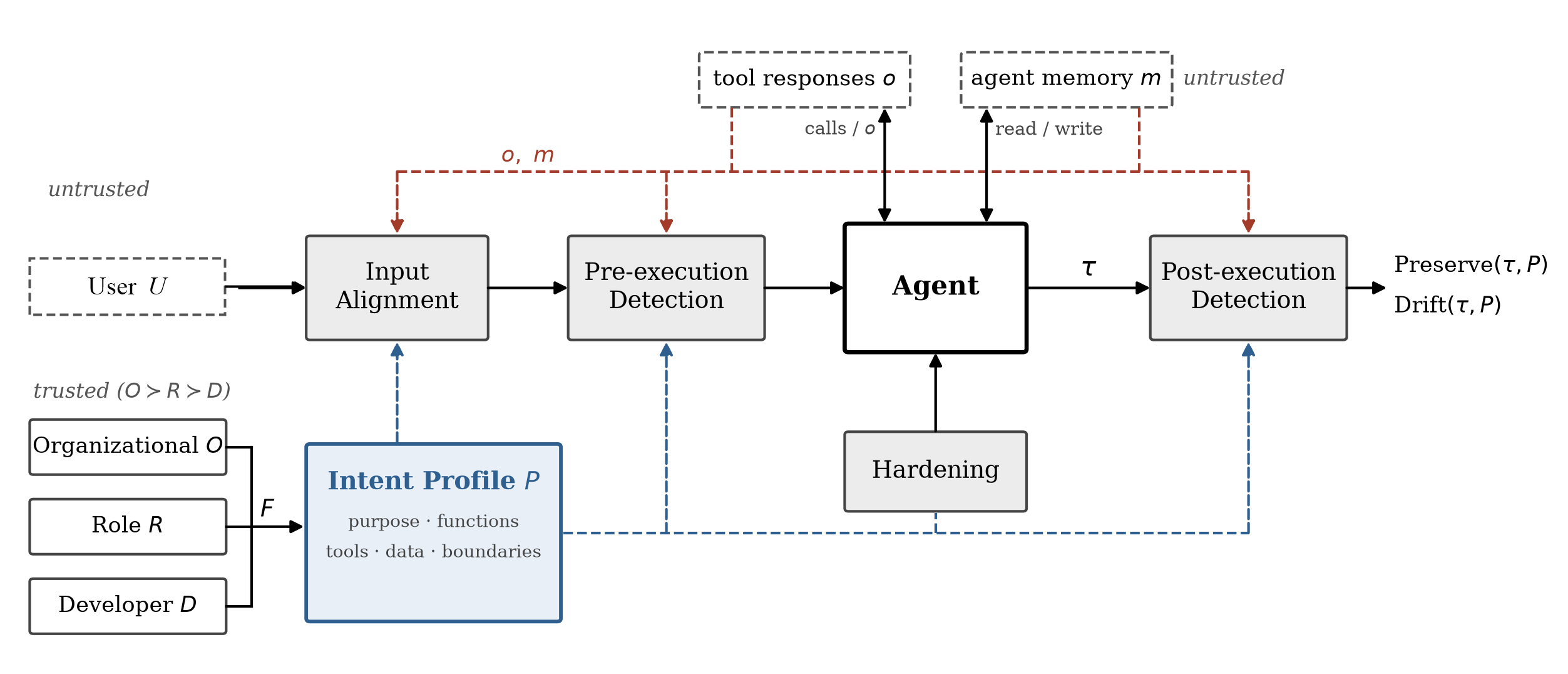}
\caption{Intent-Aware Security overview. Trusted intent sources ($O \succ R \succ D$) are extracted into an Intent Profile $P$, which conditions input alignment, pre-execution detection, hardening and post-execution detection around the agent; requests, context, tool responses and memory are untrusted.}
\label{fig:architecture}
\end{figure*}

To operationalize the agent's intent, we introduce an \textbf{Intent Profile}, an agent-specific representation derived from organizational intent, role intent, and developer intent. Organizational intent captures enterprise policy and governance constraints; role intent captures the responsibilities and operational boundaries of the deployed function; and developer intent captures the agent's configured purpose, tools, and workflows. The profile structures this information into the agent's mission, permitted tasks and workflows, tool-use constraints, data boundaries, and prohibited behaviors. The Intent Profile is not an additional source of authority. Rather, it provides security mechanisms with an explicit representation of the deployment context against which requests and behaviors can be evaluated (Fig.~\ref{fig:architecture}).

This formulation gives rise to a different security decision than conventional attack detection. Rather than asking only whether an interaction resembles an attack, Intent-Aware security asks how the interaction relates to the purpose and boundaries of the deployed agent. We operationalize this distinction using four interaction classes: \textbf{Benign}, representing legitimate use; \textbf{Out-of-Scope}, representing non-adversarial requests outside the agent's intended function; \textbf{Abusive}, representing deliberate attempts to repurpose the agent without an explicit security-compromise objective; and \textbf{Malicious}, representing interactions with an explicit attack objective. These categories allow us to evaluate whether knowledge of the agent's intent helps distinguish behaviors that may be difficult to separate from interaction content alone.

Evaluating this question requires a benchmark in which security ground truth depends on the agent receiving the interaction. We therefore introduce a generative Intent-Aware benchmark spanning heterogeneous agent configurations, domains, difficulty levels, and interaction contexts. Each sample pairs an interaction with organizational intent, role intent, and agent configuration and assigns one of the four intent-aware labels. The benchmark generation procedure can produce additional agent configurations, interactions, and contextual variations, enabling evaluation beyond a fixed collection of attack strings.

We evaluate Intent-Aware Security in three experiments. In pre-execution four-class classification on the Intent-Aware benchmark, an Intent Profile raises macro-F1 from 45\% with raw agent context to 79\%, with Out-of-Scope F1 increasing from 8\% to 76\%. On more than 10,000 executed AgentDojo trajectories with environment-derived labels, profile-based post-execution detection reaches 0.949 AUROC, compared with 0.897 for raw agent configuration and 0.909 for AlignmentCheck. Finally, profile-based Input Alignment improves protection success from 78\% to 85\% over DataFilter. Together, the results show that an explicit, agent-specific representation of intent improves fine-grained security decisions beyond raw context alone. Intent-Aware protection complements attack-focused defenses by assessing whether otherwise permissible behavior fits the agent's deployed purpose.

This paper makes three major contributions:

\begin{enumerate}
\item \textbf{Intent Preservation and Intent Profiles.} We formulate agent security in terms of preserving trusted intent governing an agent's deployment and introduce the Intent Profile, an explicit structured representation of organizational, role, and developer constraints for security decision-making.
\item \textbf{An intent-aware benchmark.} We introduce a generative Intent-Aware benchmark that evaluates interactions relative to the agent's intent of specific agents, distinguishing legitimate use, non-adversarial scope violations, deliberate agent repurposing, and explicit attacks.
\item \textbf{Intent-Aware protections.} We introduce four Intent Profile-based protection mechanisms that apply trusted agent intent to pre-execution detection, post-execution detection, input alignment, and system-prompt hardening. We evaluate their effectiveness against corresponding profile-free baselines.
\end{enumerate}

\section{Background and Related Work}\label{sec:background-and-related-work}

\subsection{The Agentic Attack Surface}\label{sec:the-agentic-attack-surface}

LLMs are increasingly embedded in agentic systems that combine model reasoning with retrieval, memory, tools, and multi-step execution --- expanding the security boundary beyond the model, since an agent interprets developer and user instructions while processing untrusted content from tools, documents, webpages, and other agents. Indirect prompt injection exploits exactly this instruction--data ambiguity, allowing an adversary to place instructions in external content that replaces or modifies the agent's original task, manipulating application behavior, triggering unauthorized tool use, and exfiltrating data~\cite{ref1}. The vulnerability generalizes to tool-integrated agents and broader agentic settings~\cite{ref2,ref3}, and surveys of agentic AI security emphasize threats arising from multi-step execution, untrusted entities, memory, and dynamic environments~\cite{ref4,ref5}. Practitioner taxonomies follow the same framing: The OWASP Top 10 for LLM Applications identifies prompt injection as a principal risk~\cite{ref6}, while the OWASP Top 10 for Agentic Applications separately addresses agent-specific risks such as goal hijacking and tool misuse~\cite{ref7}. MITRE ATLAS~\cite{ref8} catalogs adversarial techniques against AI-enabled systems.

Framing security as classification of adversarial inputs or unsafe actions is necessary but incomplete. An action can be individually benign, free of recognizable attack indicators, and still inappropriate for the purpose the agent was developed and deployed to serve.

\subsection{Prompt-Injection Defenses}\label{sec:prompt-injection-defenses}

One defense family establishes or reinforces an instruction--data boundary through prompt construction, structured input channels, instruction privilege, or model training. Spotlighting acts at prompt construction, transforming untrusted content so it cannot present itself as authoritative instruction~\cite{ref9}. Three act at training time: StruQ combines a secure front end that separates instructions and data into structured channels with a specially trained LLM that is fine-tuned to follow instructions only from the designated prompt channel~\cite{ref10}, Instruction Hierarchy trains models to prioritize higher-privilege instructions~\cite{ref11}, and SecAlign uses preference optimization to favor the authorized task over injected instructions~\cite{ref12}. In all four, the protected object is the supplied prompt, instruction channel, or privilege ordering all artifacts of the current request, not a persistent model of what the agent was deployed to do.

A second family operates at the input boundary, transforming or filtering untrusted content before it reaches the protected model. BIPIA introduced boundary-awareness and explicit-reminder defenses for indirect injection~\cite{ref13}. PIGuard is a detection-based guardrail trained with a Mitigating-Over-defense for Free objective and released with NotInject, a set of benign prompts carrying injection trigger words to measure over-defense~\cite{ref14}. It is relevant because it targets false positives, yet it classifies text as injection or not, and its over-defense is measured against benign prompts rather than against the scope of a particular agent. Polymorphic Prompt Assembling (PPA)~\cite{ref15} takes a structural route, dynamically varying the separators and assembly structure used to isolate untrusted input, making the prompt format and instruction-data boundary difficult for an attacker to predict. DataFilter is a test-time, model-agnostic sanitizer that conditions on the legitimate user instruction and the external data to strip injected content while preserving benign information~\cite{ref16}. It is the closest to one component of our approach, since it uses task context to decide what survives, but its reference is the single supplied user instruction, and it neither constructs a persistent profile of developer, role, and organizational objectives nor detects deviation after the model has reasoned or acted.

A third family moves enforcement outside the LLM: CaMeL constrains how untrusted data influences execution using control- and data-flow extracted from a trusted query~\cite{ref17}, Fides tracks confidentiality and integrity labels in agent planners~\cite{ref18}, permissive information-flow analysis reduces taint-propagation utility cost~\cite{ref19}, and AgentSpec specifies runtime behavioral constraints~\cite{ref20}. Their guarantees are expressed over predefined flow, capability, or policy specifications; we ask where that specification comes from.

\subsection{Task Drift, Goal Drift, and Contextual Security}\label{sec:task-drift-goal-drift-and-contextual-security}

Several methods locate the signal inside the model: Attention Tracker detects a \emph{distraction effect} in which attention heads shift from the original instruction to an injected one~\cite{ref21}, HiddenDetect reads safety signals from hidden-state activations~\cite{ref22}, and TaskTracker defines \emph{task drift} as deviation after untrusted data is consumed, detecting it from activation deltas with linear probes that generalize to unseen attacks~\cite{ref23}. These shift detection from injection strings toward departure from an objective, but share two deployment assumptions we do not make: white-box access to activations or attention, which excludes hosted and third-party models, and a reference objective that is the supplied task rather than a persistent role or organizational scope.

Black-box monitors pursue the same objective without activation access. LlamaFirewall's AlignmentCheck audits an agent's reasoning and actions for goal hijacking~\cite{ref43}; Task Shield verifies that each instruction and tool call contributes to the user's goals~\cite{ref44}; and MELON flags an attack when the agent's tool calls remain unchanged after the user prompt is masked~\cite{ref45}. These monitors evaluate behavior against the current user request. Our post-execution detector instead evaluates behavior against a persistent Intent Profile, enabling it to flag actions that serve the user's request but exceed the agent's deployed purpose.

A parallel line studies drift without an attacker: goal adherence degrades under environmental pressure over long trajectories~\cite{ref24} and can be inherited from other agents' trajectories~\cite{ref25}. Both ask whether a specified goal remains stable, not where it comes from.

Siu et al.~\cite{ref26} formalize contextual agent security through task alignment, action alignment, source authorization, and data isolation. Complementarily, Tsai and Bagdasarian propose Conseca, a framework for generating just-in-time, contextual, and human-verifiable security policies for an agent's current purpose~\cite{ref27}. Conseca establishes purpose-sensitive policy generation as an important direction for agent security. Our work addresses a different construction and evaluation problem: extracting an agent-specific Intent Profile from organizational, role, and developer intent, and measuring whether that representation improves fine-grained classification and protection across agents.

\subsection{Benchmarks and Alignment}\label{sec:benchmarks-and-alignment}

Existing suites evaluate agent security along several axes: BIPIA covers indirect injection in LLM applications~\cite{ref13}, InjecAgent tool-integrated agents~\cite{ref2}, AgentDojo joint security--utility scoring in an executable environment~\cite{ref28}, Agent Security Bench injection, memory poisoning, backdoors, and tool use~\cite{ref3}, and WASP end-to-end injection against web agents~\cite{ref29}. Two extend evaluation beyond attack success: AgentLure~\cite{ref30}, released with ARGUS, scores context-dependent tasks under context-aware injection, where the correct action depends on runtime evidence; StakeBench distinguishes harms by affected stakeholder and evaluates execution process alongside outcome~\cite{ref31}.

The alignment literature targets a related but distinct layer: RLHF and instruction tuning~\cite{ref32} and Constitutional AI~\cite{ref33} shape a model's behavioral prior, which degrades under deployment pressure~\cite{ref34}, while Guaranteed Safe AI~\cite{ref35} motivates specifications enforced independently of model objectives. None provides a deployment-time test of whether a given action lies within a specific agent's intended scope.

Prior work supplies the components: instruction provenance and hierarchy, injection detection and alignment, information-flow and capability enforcement, runtime policy, drift detection, contextual definitions, and attack-oriented benchmarks. Intent-Aware Security asks a question complementary to \emph{"does this interaction resemble an attack?"} to \emph{"is this behavior consistent with what this agent is supposed to do?"}

Our contribution is the joint formulation of the agent's intent as a security object, its operationalization through extraction of an agent-specific Intent Profile, an Intent-Aware benchmark for evaluating intent-conditioned security decisions, and four Intent Profile-based protection mechanisms that apply the profile to hardening, pre-execution detection, post-execution detection, and input alignment.

Our protection mechanisms differ from these prior defenses in the reference object used to make protection decisions. Existing hardening mechanisms primarily strengthen instruction-data separation or instruction privilege; detection mechanisms primarily identify attack characteristics in interaction content; and input-transformation mechanisms typically preserve a supplied task while removing adversarial content. Our mechanisms instead condition each protection point on an agent-specific Intent Profile derived from organizational, role, and developer intent, enabling protection decisions to depend on whether otherwise permissible behavior is appropriate for the specific deployed agent.

\section{Intent Preservation}\label{sec:intent-preservation}

We introduce Intent Preservation as an agent-specific security property: an agent preserves intent when its behavior remains consistent with the trusted purpose and operational constraints governing its deployment. Our formulation distinguishes the sources from which trusted agent intent is derived, the structured Intent Profile used to represent it, and the user objective evaluated against that profile.

\subsection{Trusted Intent Sources and User Intent}\label{sec:trusted-intent-sources-and-user-intent}

Let the trusted intent sources of a deployed agent be:

\[Z = \left( O,R,D \right),
\]
where \(O\), \(R\), and \(D\) denote organizational, role, and developer intent, respectively.

\textbf{Organizational intent} captures enterprise policies, governance requirements, compliance obligations, security constraints, and acceptable-use rules governing the environment in which the agent operates.

\textbf{Role intent} captures the responsibilities, workflows, escalation paths, and operational boundaries associated with the agent's deployed organizational function.

\textbf{Developer intent} captures the purpose and behavior embodied in the agent's implementation and configuration, including its system instructions, tools, knowledge sources, and expected workflows.

These sources may overlap or conflict. In our model, conflicts are resolved using the precedence relation:

\[O \succ R \succ D.
\]
A lower-priority source may refine a higher-priority source when the latter leaves behavior unspecified, but it cannot authorize behavior prohibited by a higher-priority constraint.

\textbf{User intent}, denoted \(U\), represents the objective expressed by the current user request and its interaction context. Unlike \(O\), \(R\), and \(D\), user intent is not part of the trusted specification. It may be legitimate, mistaken, out of scope, deliberately repurposing, or malicious. It is therefore evaluated against the trusted specification rather than incorporated into the security objective being preserved.

We assume that the organizational policies, role specifications, and agent configurations supplied to the extraction process are authentic and uncompromised. This assumption does not guarantee that these sources are complete, mutually consistent, or correctly represented by the extracted profile.

\subsection{Intent Profile}\label{sec:intent-profile}

The trusted specification may be distributed across heterogeneous policies and agent configurations. We therefore define an extraction function:

\[E:Z \rightarrow P,
\]
which converts the trusted specification \(Z\) into an agent-specific Intent Profile:

\[P = \left( P_{\mathrm{purpose}},P_{\mathrm{function}},P_{\mathrm{tool}},P_{\mathrm{data}},P_{\mathrm{boundary}} \right).
\]

The profile contains five dimensions:

\begin{itemize}
\item \(P_{\mathrm{purpose}}\): the agent's mission, expected outcomes, and success criteria.
\item \(P_{\mathrm{function}}\): the workflows, task categories, and responsibilities the agent may perform.
\item \(P_{\mathrm{tool}}\): the tools the agent may use and the conditions, operations, and parameter constraints governing their use.
\item \(P_{\mathrm{data}}\): the permitted data sources, sensitivity restrictions, and purpose-bound conditions on data use.
\item \(P_{\mathrm{boundary}}\): prohibited behavior and the conditions requiring refusal, redirection, or escalation.
\end{itemize}

The Intent Profile does not introduce an additional source of authority. It structures the constraints derived from \(O\), \(R\), and \(D\) into a representation that can be used by protection mechanisms.

Because \(P\) is agent-specific, the security interpretation of an interaction cannot always be determined from the interaction alone. For example, analyzing a prompt-injection payload may be consistent with the profile of a security-analysis agent but outside the profile of a travel-planning agent.

\subsection{Intent Preservation and Intent Drift}\label{sec:intent-preservation-and-intent-drift}

Let an agent execution be represented as:

\[\tau = \left( u,h_{0},a_{1},\ldots,a_{T} \right),
\]
where \(u\) is the current user request, \(h_{t}\) is the execution context at step \(t\), and \(a_{t}\) is an observable agent behavior. Behaviors may include generated answers, plans, retrieval decisions, tool invocations, or memory modifications.

Let:

\[C\left( a_{t},h_{t},P \right) \in \left\{ 0,1 \right\}
\]
denote whether behavior \(a_{t}\), in context \(h_{t}\), is consistent with the constraints encoded by Intent Profile \(P\).

An execution preserves intent when:

\[\mathrm{Preserve}\left( \tau,P \right) \iff \forall t \in \left\{ 1,\ldots,T \right\},\ C\left( a_{t},h_{t},P \right) = 1.
\]

An execution exhibits Intent Drift when this property is violated:

\[\mathrm{Drift}\left( \tau,P \right) \iff \exists t \in \left\{ 1,\ldots,T \right\}:C\left( a_{t},h_{t},P \right) = 0.
\]
A non-benign request represents a potential cause or attempted induction of Intent Drift, not necessarily a completed drift event. Intent Drift occurs only when the agent's resulting behavior violates its Intent Profile. An agent may therefore preserve intent despite receiving an Out-of-Scope, Abusive, or Malicious request, provided that it refuses, safely redirects, or otherwise responds consistently with its Intent Profile.

The same distinction applies to indirect prompt injection. An adversarial instruction embedded in retrieved content or a tool response appears in the execution context \(h_{t}\) and constitutes an attempted induction of Intent Drift. The attempt becomes a completed drift event only if it influences an observable behavior \(a_{t}\) such that \(C\left( a_{t},h_{t},P \right) = 0\). If the agent ignores the instruction, treats it as data, refuses the induced action, or otherwise responds consistently with \(P\), the execution preserves intent despite the presence of an indirect prompt-injection attack.

Intent Preservation is stricter than capability authorization. Capability authorization asks whether an agent is permitted to invoke a tool or perform an operation. Intent Preservation additionally asks whether that behavior, in its execution context, serves a purpose consistent with the agent's mission and operational boundaries. Consequently, an authorized tool call may still violate intent when it is performed for an unintended purpose.

The predicate \(C\) represents the security judgment that Intent-Aware mechanisms seek to approximate. We do not assume that consistency can be determined perfectly. We evaluate practical approximations based on profile-aware classification, executed trajectories, and simulated behavioral outcomes.

\subsection{Request-Level Operationalization}\label{sec:request-level-operationalization}

Our formal definition applies to an execution trajectory, but protection decisions must often be made before or during execution. We therefore operationalize the relationship among a request \(u\), its context \(h\), and Intent Profile \(P\) using four intent-aware interaction classes.

Let:

\begin{itemize}
\item \(S\left( u,h,P \right)\) indicate whether the requested behavior is consistent with \(P\);
\item \(A\left( u,h \right)\) indicate the presence of an explicit attack or security-compromise objective;
\item \(D\left( u,h,P \right)\) indicate a deliberate attempt to repurpose the agent beyond \(P\).
\end{itemize}

We assign labels using the following ordered decision rule:

\begin{equation*}
Y(u,h,P) =
\begin{cases}
\text{Malicious}, & A(u,h) = 1,\\
\text{Abusive}, & A = 0 \land S = 0 \land D = 1,\\
\text{Out-of-Scope}, & A = 0 \land S = 0 \land D = 0,\\
\text{Benign}, & A = 0 \land S = 1,
\end{cases}
\end{equation*}
where $A$, $S$, and $D$ abbreviate $A(u,h)$, $S(u,h,P)$, and $D(u,h,P)$.

The ordering resolves interactions that could otherwise satisfy multiple descriptions. Explicit attack objectives take precedence over repurposing, and deliberate repurposing takes precedence over non-adversarial scope mismatch.

\begin{itemize}
\item \textbf{Benign:} The requested behavior is consistent with the agent's Intent Profile.
\item \textbf{Out-of-Scope:} A non-adversarial request for behavior inconsistent with the agent's Intent Profile.
\item \textbf{Abusive:} A deliberate attempt to repurpose the agent beyond its intended purpose, without an explicit security-compromise objective.
\item \textbf{Malicious:} An interaction involving an explicit attack objective, such as instruction override, security-control bypass, unauthorized access, extraction, or destructive action.
\end{itemize}

These labels operationalize the relationship between an interaction and the receiving agent's Intent Profile; they do not establish whether Intent Drift occurs. Benign, Out-of-Scope, Abusive, and Malicious describe the relationship between an interaction and the receiving agent's Intent Profile. Intent Preservation and Intent Drift, by contrast, are trajectory-level properties determined by the agent's resulting behavior. A non-benign interaction may therefore result in preserved intent if the agent refuses, redirects, or safely constrains it, while a completed drift event requires observable behavior inconsistent with the profile. Experiment I evaluates interaction-level classification, Experiment II detects drift in executed trajectories, and Experiment III approximates protection effects through simulated responses.

\section{Threat Model}\label{sec:threat-model}

We consider a deployed agent governed by trusted intent sources and an Intent Profile extracted from them. The adversary's goal is to induce Intent Drift: to cause the agent's behavior to depart from its Intent Profile. Following Section~\ref{sec:request-level-operationalization}, we distinguish two adversarial goals. A Malicious adversary pursues an explicit security-compromise objective, such as instruction override, data exfiltration, unauthorized access, or destructive action, corresponding to goal hijacking in the OWASP Top 10 for Agentic Applications~\cite{ref7}. An Abusive adversary seeks to repurpose the agent beyond its deployed function, for example by inducing individually authorized tool use for an unauthorized purpose, related to excessive agency in the OWASP Top 10 for LLM Applications~\cite{ref6}. Out-of-Scope requests are non-adversarial and therefore outside the threat model; we include them in evaluation because a deployed protection must distinguish them from attacks.

\paragraph{Attacker capabilities} The adversary may act as a user, submitting arbitrary requests and multi-turn conversations (direct injection)~\cite{ref6,ref26}, or as a third party controlling content the agent consumes, including emails, documents, webpages, tickets, and tool outputs (indirect injection)~\cite{ref1,ref2}. We assume the adversary knows the agent's general purpose and available tools, but cannot modify the agent's configuration, organizational or role policies, the profile-extraction process, or the protection mechanisms.

\paragraph{Trust assumptions and out-of-scope threats} Consistent with Section~\ref{sec:trusted-intent-sources-and-user-intent}, we assume trusted intent sources are authentic and uncompromised; we do not assume they are complete or consistent. Compromise of these sources, of model weights, or of the hosting infrastructure is out of scope.

Finally, we treat tool-call responses and the agent's memory as untrusted.

\section{Intent-Aware Security Benchmark}\label{sec:intent-aware-security-benchmark}

Existing agent-security benchmarks primarily evaluate whether an interaction contains an attack, causes a harmful outcome, or violates a security property. Our benchmark addresses a complementary question: \emph{does the security interpretation of an interaction depend on the purpose and operational boundaries of the agent receiving it?}

Each benchmark sample therefore evaluates a user interaction relative to trusted organizational intent, role intent, and agent configuration. This enables interactions to be classified into the four intent-aware classes defined in Section~\ref{sec:intent-preservation}: Benign, Out-of-Scope, Abusive, and Malicious. The benchmark is designed to distinguish legitimate use from non-adversarial scope violations, deliberate agent repurposing, and explicit attacks, including cases in which similar behavior has different security interpretations for different agents.

\subsection{Benchmark Construction}\label{sec:benchmark-construction}

The benchmark is generated through a hybrid procedure that combines deterministic construction with LLM-based natural-language realization. A central design goal is to separate the benchmark specification from its realization. The deterministic construction procedure fixes each sample's agent, action, target label, difficulty, and, where applicable, attack family and delivery mode before any natural-language content is generated. The benchmark therefore does not ask an LLM to freely generate an interaction and subsequently decide what security label it represents. Instead, an LLM realizes the predetermined coordinate as a concrete agent configuration and interaction.

Benchmark construction therefore consists of two types of operations.

\textbf{Deterministic construction.} The construction procedure determines the agent-design grid, selects actions, assigns target labels, assigns difficulty levels, schedules attack families and delivery modes, and determines organizational/role-policy alignment conditions. These decisions define the intended semantic properties of a sample before the interaction is generated.

\textbf{LLM-based realization.} Given these predetermined conditions, sample-realization-LLM generates the natural-language artifacts required to instantiate them. These artifacts include organizational and role policies; agent configurations, including system prompts, tools, and knowledge-source descriptions; and interaction content, including the current user request, conversation history, and tool responses.

Consequently, sample-realization-LLM does not assign the benchmark ground-truth label. The target label is established by the benchmark deterministic construction procedure before natural-language generation.

\subsection{Agent Design Space}\label{sec:agent-design-space}

Agents are generated from a structured design grid intended to produce heterogeneous purposes and authorization boundaries. Each agent is defined along five dimensions:

\begin{enumerate}
\item \textbf{Domain:} the organizational or operational area in which the agent operates.
\item \textbf{Pattern:} the architectural or orchestration pattern used by the agent.
\item \textbf{Purpose:} the outcome the agent is intended to achieve.
\item \textbf{Workflow:} the business process or activity supported by the agent.
\item \textbf{Capability:} an action-resource-scope tuple describing an available operation and its authorized scope.
\end{enumerate}

The design procedure emphasizes intent diversity rather than domain diversity alone. Agents are selected so that meaningful actions can have different authorization states across profiles. For example, an action that is a legitimate responsibility for one agent may be outside the role or purpose of another. This allows the benchmark to test whether a security mechanism reasons about the relationship between an interaction and the deployed agent rather than only recognizing properties of the interaction itself.

For each selected design coordinate, sample-realization-LLM instantiates the corresponding organizational policy, role policy, system prompt, available tools, and knowledge-source descriptions. Agent configurations may either align with the higher-level organizational and role policies or omit constraints established by those policies. This allows the benchmark to represent cases in which the immediate agent configuration is not sufficient on its own to reconstruct the complete Intent Profile.

\subsection{Intent-aware Label Assignment}\label{sec:intent-aware-label-assignment}

Each interaction is assigned a target label Benign, Out-of-scope, Abusive, or Malicious according to the definition in Section~\ref{sec:request-level-operationalization}. Importantly, labels are assigned at the profile-action level before conversation realization. The sample-realization-LLM receives the predetermined construction coordinate and generates an interaction intended to instantiate it. The subsequent validation stage determines whether the generated sample actually provides sufficient evidence for the assigned class. Samples that fail this requirement are rejected rather than having their labels changed.

This construction also permits adversarial-looking content to occur in non-Malicious samples. For example, an agent whose legitimate purpose is security analysis may be asked to analyze prompt-injection content as data. Conversely, an interaction containing no recognizable attack string may still be Out-of-Scope or an Abusive when the requested behavior conflicts with the receiving agent's purpose.

\subsection{Interaction and Challenge Generation}\label{sec:interaction-and-challenge-generation}

Once an agent-action-label coordinate is established, the benchmark construction procedure deterministically assigns the challenge level, delivery mode, and, for Malicious samples, attack family and delivery mode before generation. The sample-realization-LLM then realizes these predetermined conditions as a concrete interaction containing the current user request and, when required by the assigned delivery mode, historical conversation and tool responses.

Samples span three predetermined challenge levels: Easy, Mid, and High. Easy samples represent relatively direct interactions with limited contextual ambiguity. Mid- and High-difficulty samples progressively introduce assigned challenge factors, including multilingual or encoded content, instruction-like text used as data, contextual ambiguity, conflicting instructions, false authority, and multiple information sources. The sample-realization-LLM instantiates these assigned conditions but does not select the sample's challenge level.

For Malicious samples, the attack family and delivery mode are likewise assigned before interaction generation. Direct attacks place the adversarial instruction in user-controlled interaction content, whereas indirect attacks place it in untrusted context such as a tool response. Attack families are derived from public attack datasets. The sample-realization-LLM instantiates the assigned attack family and delivery mode within the agent-specific benchmark context.

This construction separates benchmark design decisions from natural-language realization: the construction procedure determines the target label, challenge level, delivery mode, and applicable attack conditions, while the sample-realization-LLM generates an interaction satisfying those predetermined conditions.

\subsection{Benchmark Validation}\label{sec:benchmark-validation}

Because the benchmark uses LLMs to realize synthetic agents and interactions, generated samples undergo layered validation before inclusion. Deterministic validation checks that each sample satisfies the required schema, label, challenge, delivery-mode, metadata, and content constraints. Candidates then undergo independent semantic validation by a realization-judge-LLM distinct from the sample-realization model. The judge evaluates construction-plan consistency, label correctness, intent fit, attack-technique fit, and difficulty; a sample is accepted only if every criterion scores at least 4 on a five-point scale.

To reduce judge variance, we run semantic validation twice and retain the lower score. Where the judge provides supporting evidence, we verify it deterministically and incorporate the result into the final validation score. Finally, to independently audit the validation pipeline, we manually reviewed a random subset of accepted samples.

\section{Intent-Aware Protections}\label{sec:intent-aware-protections}

The preceding sections define the Intent Profile as a structured representation of an agent's mission, permitted workflows, tool constraints, data boundaries, and prohibited behaviors. We next ask whether this representation is useful when incorporated into different types of agent protection.

We introduce four Intent Profile-based protection mechanisms that operationalize trusted agent intent at complementary points in the agent pipeline:

\begin{enumerate}
\item \textbf{Intent Profile-based Pre-execution Detection}, which uses the profile to classify an interaction \emph{before} execution.
\item \textbf{Intent Profile-based Post-execution Detection}, which uses the profile to classify an interaction \emph{after} execution.
\item \textbf{Intent Profile-based Input Alignment}, which uses the profile to preserve, transform, or block interaction content before it reaches the agent.
\item \textbf{Intent Profile-based Hardening}, which incorporates the profile into the agent's system instructions.
\end{enumerate}

The novelty of these mechanisms lies in conditioning protection decisions on an agent-specific Intent Profile derived from organizational, role, and developer intent, rather than on the interaction, system prompt, or current task alone.

\subsection{Intent Profile Extraction}\label{sec:intent-profile-extraction}

To evaluate the contribution of an Intent Profile across the protection mechanisms, we first extract a structured profile from three trusted sources: organizational intent, role intent, and developer intent.

For this evaluation, we represent developer intent through an agent configuration schema that captures the agent's system instructions, declared tools, knowledge sources, workflows, capabilities, and relevant memory configuration. For simplicity, we represent organizational intent and role intent as unstructured text fields summarizing the policies, responsibilities, constraints, and operational boundaries applicable to the agent. Real organizational and role intent would typically be distributed across multiple policy documents, governance systems, access-control configurations, operating procedures, and other organizational artifacts. The text summaries used in this evaluation are therefore an experimental abstraction rather than a complete representation of a production environment.

Together, these choices allow the extraction task to cover two input forms:

\begin{enumerate}
\item \textbf{Semi-structured input}, represented by the low-code-style agent configuration that encodes developer intent.
\item \textbf{Unstructured input}, represented by the textual organizational-intent and role-intent summaries.
\end{enumerate}

The extraction process first analyzes the agent configuration to identify its stated purpose, supported workflows, declared capabilities, available tools, knowledge sources, data boundaries, and explicit behavioral constraints. It then incorporates the organizational-intent and role-intent summaries and resolves overlapping or conflicting constraints according to the precedence relation defined in Section~\ref{sec:intent-preservation}.

The extracted profile is held constant across the protection conditions described below. This enables the experiments to evaluate how the same explicit representation of the agent's intent contributes when applied to pre- and post-execution detection, input alignment, and hardening.

\subsection{Intent Profile-Based pre-execution Detection}\label{sec:intent-profile-based-pre-execution-detection}

The pre-execution detection tests whether the Intent Profile can support a pre-execution decision about the relationship between an interaction and the purpose of the deployed agent.

The detector receives: (1) the Intent Profile; (2) the current user request; and (3) available interaction context, including conversation history and tool responses. It assigns one of the four intent-aware labels defined in Section~\ref{sec:intent-preservation}: (1) Benign; (2) Out-of-Scope; (3) Abusive; or (4) Malicious.

The detector evaluates the interaction relative to the receiving agent. It therefore differs from a conventional prompt-injection detector whose primary objective is to recognize adversarial patterns in the interaction itself. An interaction can contain suspicious or instruction-like text while remaining legitimate for a security-analysis agent. Conversely, a request can contain no recognizable attack-pattern while attempting to repurpose an agent beyond its deployed function.

This condition asks: Does conditioning a pre-execution detector on the Intent Profile improve its ability to distinguish legitimate use, scope violations, deliberate repurposing, and malicious interactions?

\subsection{Intent Profile-Based post-execution Detection}\label{sec:intent-profile-based-post-execution-detection}

The post-execution detection tests whether the Intent Profile can support a post-execution decision about the relationship between an agent execution trajectory and the purpose of the deployed agent.

The detector receives: (1) the Intent Profile; (2) the current user request; (3) available execution context, including ordered tool calls, tool output, intermediate messages, environment changes, and (4) final answer. It assigns one of the four intent-aware labels defined in Section~\ref{sec:intent-preservation}: (1) Benign; (2) Out-of-Scope; (3) Abusive; or (4) Malicious.

The detector evaluates the agent's resulting behavior relative to the receiving agent's Intent Profile, rather than the interaction content alone. It therefore approximates the trajectory-level consistency predicate defined in Section~\ref{sec:intent-preservation-and-intent-drift}: Intent Drift occurs only when observable behavior violates the profile. A trajectory may contain an injected or instruction-like payload yet preserve intent if the agent ignores, refuses, or safely constrains it. Conversely, a trajectory may contain no recognizable attack pattern while the agent invokes individually authorized tools for a purpose outside its deployed function.

This condition asks: Does conditioning a post-execution detector on the Intent Profile improve its ability to distinguish executions that preserve the agent's intent from executions in which the agent's behavior has drifted beyond its deployed purpose?

\subsection{Intent Profile-based Input Alignment}\label{sec:intent-profile-based-input-alignment}

The alignment condition tests whether the Intent Profile can be used to transform interaction content into a form that remains consistent with the agent's purpose and constraints.

User requests and retrieved content may be ambiguous, underspecified, partially in scope, or mixed with instructions that conflict with the agent's role.

The alignment mechanism considers the current request, conversation history, and tool responses. It may perform:

\begin{itemize}
\item Intent interpretation and disambiguation;
\item Removal or isolation of instructions originating from untrusted content;
\item Normalization of a request to an authorized workflow;
\item Restriction of requested tools, data, operations, or scope;
\item Blocking when no profile-consistent formulation exists.
\end{itemize}

The mechanism is also applied to conversation history and tool responses. This allows the evaluation to include interactions in which conflicting instructions appear indirectly in retrieved documents, previous messages, or tool-provided content.

This condition asks: \emph{Can an Intent Profile guide content transformation that reduces fulfillment of unauthorized objectives while preserving legitimate user intent?}

Unlike detection, alignment does not primarily assign a security category. Its output is the content presented to the agent or a decision that no profile-consistent input exists.

\subsection{Intent Profile-based Hardening}\label{sec:intent-profile-based-hardening}

The Profile-based hardening condition tests whether the Intent Profile can improve the instructions governing the agent before interaction processing begins.

Many agent system prompts describe the agent's expected functionality but leave operational boundaries implicit. They may identify the agent as an HR assistant, travel planner, or security analyst without specifying which workflows, tools, data, or actions are outside its authorized purpose. Such omissions make it difficult for the agent to distinguish a legitimate variation of its task from a request that attempts to broaden or replace that task.

In Profile-based hardening, the original system prompt and the Intent Profile are used to produce a profile-grounded system prompt. The hardened prompt preserves the agent's intended functionality while making the profile's constraints explicit.

The hardened prompt is then used as the effective system instruction in the agent-response simulation. The user request, conversation history, tool responses, and declared tools otherwise remain unchanged.

\subsection{Evaluation Scope}\label{sec:evaluation-scope}

The protection mechanisms evaluated here are operational approximations of the Intent Preservation property defined in Section~\ref{sec:intent-preservation}. Pre-execution detection, input alignment, and hardening act before execution, whereas post-execution detection evaluates an observed execution trajectory. Although post-execution detection receives the complete recorded agent trajectory, its monitor is an empirical approximation rather than a formal verifier of every execution step and therefore does not establish Intent Preservation as a guarantee.

Accordingly, our experiment addresses the following question:

\emph{Does an explicit representation of the agent's intent provide useful conditioning information across pre-execution detection, post-execution detection, hardening, and input alignment?}

\section{Experimentation and Results}\label{sec:experimentation-and-results}

We evaluate Intent-Aware protections through three tracks: pre-execution classification on our benchmark (Exp. I), post-execution detection on executed AgentDojo~\cite{ref28} trajectories (Exp. II), and protection intervention on our benchmark (Exp. III). Classification experiments measure the agreement between detector predictions and four-class benchmark ground truth, whereas protection experiments measure the downstream behavioral effect of applying a protection. Consequently, detector accuracy and protection effectiveness are reported and analyzed separately.

\subsection{Benchmark Dataset and Experimental Unit}\label{sec:benchmark-dataset-and-experimental-unit}

We configure Malicious cases based on Garak~\cite{ref36} attack families, and they are configured to include both direct and indirect attack delivery. The benchmark uses GPT-5.4~\cite{ref37} as a sample-realization-LLM, and MAI-Thinking-1~\cite{ref38} as the realization-judge-LLM. In addition, we configure the benchmark-generation to produce 1,000 samples across four intent-aware classes: Benign, Out-of-Scope, Abusive, and Malicious, with 25\% of samples assigned to each class. Challenge levels comprise 15\% Easy, 45\% Mid, and 40\% High.

Of the 1,000 generated candidates, 11 failed deterministic structural and content validation (Section~\ref{sec:benchmark-validation}), and a further 102 were rejected by the realization-judge-LLM during semantic validation. This left 887 accepted samples, an overall acceptance rate of 88.7\%. The final benchmark contains 887 accepted samples spanning 35 agent configurations. By class, the benchmark contains 237 Benign, 239 Out-of-Scope, 226 Abusive, and 185 Malicious samples. By challenge level, it contains 140 Easy, 399 Mid, and 348 High samples. The 185 Malicious samples span 12 attack families, with 104 delivered directly and 81 indirectly. The final evaluation set contains 887 accepted samples and is held fixed across all reported experimental conditions.

A random subset of 100 passing samples was then reviewed manually, and 98\% were found correctly labeled.

\subsection{Dataset Selection and Preprocessing}\label{sec:dataset-selection-and-preprocessing}

Experiments use the labeled benchmark. A benchmark sample is included only if it passed benchmark evaluation. Each accepted sample is normalized into the following experimental representation: (1) a sample identifier; (2) the current user request (user\_prompt); (3) conversation memory, including previous conversations and tool responses where available; (4) organization intent; (5) role intent; (6) agent configuration, including the system prompt and declared tools; (7) and benchmark label.

Thus, the sample does not include any information that was used to design the agent, the content, or the label. This differs substantially from the information that the generator LLM used to generate the samples' content.

The experimental unit is a single labeled sample. Multiple samples can correspond to the same underlying agent. Intent-profiling is performed per agent and uses the organization intent, role intent, and agent configuration, including the system prompt and declared tools. Input classification, response simulation and evaluation remain sample-level operations.

Each sample carries one of the four benchmark ground-truth labels defined in Section~\ref{sec:request-level-operationalization} --- Benign, Out-of-Scope, Abusive, and Malicious --- which is used only for reporting and is withheld from all detectors, simulators, and judges.

\subsection{Experiment I: Profile-based pre-execution Detection}\label{sec:experiment-i-profile-based-pre-execution-detection}

The interaction classification measures the ability of different detectors to classify benchmark conversations according to their relationship to the receiving agent's Intent Profile.

We compare four primary approaches: (1) An Input-only LLM that receives the untrusted input only without knowledge of the agent's configuration and Intent Profile; (2) PIGuard~\cite{ref14}, a specialized profile-free prompt-injection classifier receives untrusted input only; (3) Raw-context LLM: an LLM classifier that receives the untrusted input and agent configuration, organizational intent, role intent, without knowledge the extracted Intent Profile; and (4) Intent Profile-based Pre-execution Detection: an LLM classifier that receives the untrusted input and the extracted Intent Profile.

We compare four methods under the binary Malicious versus Non-Malicious evaluation. For the primary four-class evaluation, we compare the three LLM classifiers capable of producing the benchmark's four labels.

We use DeepSeek-V4-Pro~\cite{ref39} for LLM-based classification with the same configuration across all detectors. All detectors receive the user request, conversation history, and tool responses as input, and differ only in the agent-specific security context supplied to the classifier and how that context is represented. We chose DeepSeek-V4-Pro to rule out same-model-family effects between benchmark generation and classification.

PIGuard is loaded locally using a text-classification pipeline with input truncation at 512 tokens. Because some conversations exceed PIGuard's input length, we split each sample into overlapping 450-token chunks with a 50-token overlap and classify all chunks. A sample is labeled as injection if any chunk is classified as injection, ensuring that signals appearing anywhere in long conversations are retained.

\paragraph{Classification Metrics}\label{sec:classification-metrics}

The classifiers produce predictions over the benchmark's four intent-aware classes: Benign, Out-of-Scope, Abusive, and Malicious. We therefore use per-class F1 and macro-F1 over the four-class taxonomy as the primary evaluation. This setting directly measures whether a classifier can distinguish legitimate use, non-adversarial scope violations, deliberate agent repurposing, and explicit attacks relative to the deployed agent's intent.

The four-class evaluation remains our primary experiment because it directly tests our central research question: whether conditioning security decisions on the agent's intent improves the ability to distinguish legitimate use, scope violations, agent repurposing, and attacks.

\paragraph{Statistical analysis} Because multiple benchmark samples correspond to the same underlying agent, we use a paired cluster bootstrap analysis at the agent level, and recompute macro-F1 for both classification conditions on the same resampled agents. We report the 2.5th and 97.5th percentiles of this distribution as the 95\% bootstrap confidence interval.

\paragraph{Results}\label{sec:results-expI}

As a secondary comparison, we collapse the taxonomy into Malicious versus Non-Malicious. Under this conventional binary task, Intent Profile-based Pre-execution Detection achieves 72\% F1, compared with 50\% for Raw-context, 41\% for Input-only, and 36.8\% for PIGuard. The primary four-class evaluation below tests the stronger question of whether a detector can distinguish attacks from non-adversarial scope violations and deliberate repurposing.

We next present the classifiers' results under the full four-class taxonomy. The Intent Profile provides explicit context for determining not only whether an interaction is unsafe, but how it relates to the purpose and boundaries of the deployed agent. Results are shown in Table~\ref{tab:exp1}.

\begin{table}[t]
\centering
\caption{Four-class pre-execution classification on the Intent-Aware benchmark (F1).}
\label{tab:exp1}
\small
\setlength{\tabcolsep}{3pt}
\begin{tabular}{@{}lccccc@{}}
\toprule
\textbf{Method} & \textbf{Macro} & \textbf{Mal.} & \textbf{Abus.} & \textbf{OoS} & \textbf{Benign} \\
\midrule
Input-only & 43\% & 41\% & 70\% & 2\% & 60\% \\
Raw-context & 45\% & 50\% & 57\% & 8\% & 64\% \\
Profile-based & \textbf{79\%} & \textbf{72\%} & \textbf{86\%} & \textbf{76\%} & \textbf{82\%} \\
\bottomrule
\end{tabular}
\end{table}

The Input-only LLM, which receives no agent-specific intent context, achieved a macro-F1 of 43\%. The Raw-context LLM that uses the same classifier with the underlying organizational intent, role intent, and agent configuration without extracting an Intent Profile, achieved a macro-F1 of 45\%. However, the profile-based classifier further increased macro-F1 to 79\%. Thus, access to raw agent context accounts for two percentage-point improvement over the baseline, while structuring this context into an explicit Intent Profile provides an additional 34 percentage-point improvement. The profile-based classifier also achieved the highest F1 across all four classes, with particularly large gains for Out-of-Scope interactions.

Under paired agent-level bootstrap resampling, the 95\% confidence interval for the 34-percentage-point macro-F1 improvement of Intent Profile-based Detection over the Raw-context LLM was {[}29.5, 39.0{]} percentage points. This indicates that the profile-based advantage persists under variation in the composition of agent configurations rather than being driven solely by particular agents in the benchmark.

The profile's primary advantage emerges in the four-class task, where the explicit Intent Profile enables the classifier to distinguish how an interaction relates to the agent's intended purpose and operational boundaries, rather than merely separating Malicious from Non-Malicious interactions.

Appendix~\ref{app:confusion} shows the confusion-difference matrix relative to the Raw-context classifier. The largest recall gains are for Out-of-Scope (+74.0 pp), mostly from samples previously labeled Benign (--56.1 pp), and for Malicious (+41.6 pp), mostly from samples previously labeled Abusive (--41.5 pp). Abusive recall decreases by 15.9 pp, but far fewer Benign, Out-of-Scope, and Malicious samples are mislabeled as Abusive, so Abusive precision improves and Abusive F1 rises from 57\% to 86\%.

\subsection{Experiment II: Intent Profile-based Post-execution Detection}\label{sec:experiment-ii-intent-profile-based-post-execution-detection}

The post-execution detection evaluation measures whether the Intent Profile improves detection of Intent Drift from completed agent executions. Unlike Experiments I and III, this evaluation does not use our benchmark; trajectories are produced by simulation of tool execution and labeled by deterministic environment checks. We use AgentDojo~\cite{ref28} as an external benchmark for this experiment. Although its threat model differs from ours in that it treats user input as trusted, it remains a useful point of comparison for evaluating the full agent trajectory. We use AgentDojo for its ability to deterministically label successful attacks in its simulated environment.

We use the full AgentDojo v1.2.2 benchmark, comprising four tool-using agents: Workspace (email, calendar, cloud files), Travel (flights, hotels, trip planning), Banking (accounts and transactions), and Slack (messaging and workspace administration). Attacks span eleven automated prompt-injection techniques: direct, ignore\_previous, system\_message, injecagent, tool\_knowledge, and six important\_instructions variants. Manual and denial-of-service attacks are excluded (i.e., 485 cases). All trajectories are generated without any defense using GPT-4o~\cite{ref46} as the underlying LLM. Of 11,021 configured cases, 10,536 produced usable trajectories: 97 benign runs and 10,439 attacked runs.

\textbf{Ground truth:} AgentDojo's deterministic, task-specific evaluators inspect executed tool calls, final answers, and pre- and post-execution environment state to determine whether the complete injected objective succeeded. Because AgentDojo provides neither Out-of-Scope nor Abusive interactions and labels outcomes rather than requests, we evaluate the detector of Section~\ref{sec:intent-profile-based-post-execution-detection} on a collapsed binary version of the four-class taxonomy: Malicious interactions whose resulting behavior completes the injected objective form the positive class, and all other executions, including Benign runs and Malicious interactions in which the agent preserved its intent, form the negative class. The run contains 2,480 successful attacks and 7,959 failed attacks, a 23.8\% attack-success rate. This labeling deliberately separates the \emph{presence} of an attack from its \emph{effect}: a detector must determine whether the agent's behavior drifted, not merely whether adversarial content appeared.

We compare four post-execution monitors that receive the same user request, ordered tool calls, tool outputs, environment changes, and final answer, and output a maliciousness score in {[}0, 1{]}: (1) Trajectory-only LLM, which receives the execution trajectory without agent-specific context; (2) Agent-configuration LLM, which additionally receives the raw system prompt, and agent meta-data; and (3) Intent Profile-based Post-execution Detection, which receives the trajectory and the extracted Intent Profile, comprising the agent's purpose, authorized workflows, capabilities, scope boundaries, tool-output trust boundaries, and argument-provenance constraints. (4) AlignmentCheck from LlamaFirewall~\cite{ref43}, an open-source LLM-based monitor that audits agent actions for goal hijacking. For this evaluation, we modify AlignmentCheck to return a continuous misalignment score in {[}0, 1{]} rather than its native Boolean conclusion. Unlike Intent Profile-Based Post-Execution Detection, which uses the agent's persistent Intent Profile as the authorized objective, AlignmentCheck treats the original user request as the authorized objective. Because AgentDojo also assumes a trusted user request, this setup is consistent with the benchmark and gives a fair comparison.

All monitors use GPT-4o with identical configuration. We use GPT-4o for these defenses to stay in alignment within the same attack complexity as the original AgentDojo benchmark.

\paragraph{Results}\label{sec:results-expII}

Our primary metric is AUROC over the continuous scores, because it is threshold-independent and the monitors' scores are not calibrated to a common scale. As secondary metrics, we report precision, recall, and F1 at a fixed 0.5 threshold.

\begin{table}[t]
\centering
\caption{Post-execution detection on executed AgentDojo trajectories. Precision, recall and F1 at a 0.5 threshold.}
\label{tab:exp2}
\small
\setlength{\tabcolsep}{4pt}
\begin{tabular}{@{}lcccc@{}}
\toprule
\textbf{Method} & \textbf{Prec.} & \textbf{Recall} & \textbf{F1} & \textbf{AUROC} \\
\midrule
AlignmentCheck & \textbf{95.9\%} & 91.2\% & \textbf{93.5\%} & 90.9\% \\
Trajectory-only & 65.3\% & 70.1\% & 67.6\% & 84.3\% \\
Agent-configuration & 68.4\% & 91.4\% & 78.2\% & 89.7\% \\
Profile-based & 75.9\% & \textbf{93.3\%} & 83.8\% & \textbf{94.9\%} \\
\bottomrule
\end{tabular}
\end{table}

\begin{figure}[t]
\centering
\includegraphics[width=\columnwidth]{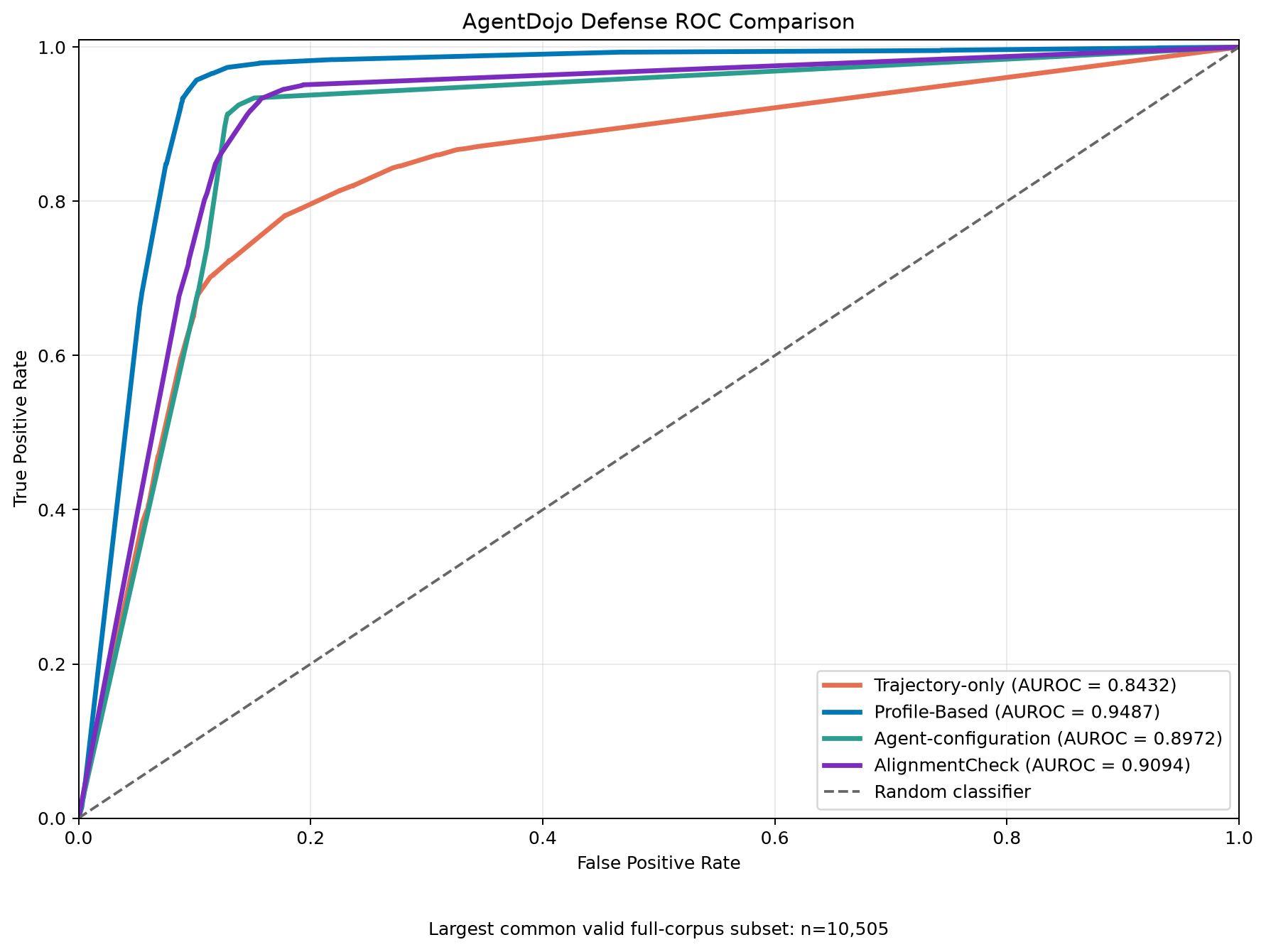}
\caption{ROC curves of the post-execution monitors on AgentDojo (largest common valid subset, $n=10{,}505$).}
\label{fig:roc}
\end{figure}

Intent Profile-based Detection achieves the highest AUROC (94.9\%): 10.6 points above Trajectory-only, 5.2 above Agent-configuration, and 4.0 above AlignmentCheck. At the fixed 0.5 threshold, AlignmentCheck attains higher precision (95.9\% vs. 75.9\%) and F1 (93.5\% vs. 83.8\%), while the profile-based monitor attains the highest recall (93.3\%). Fig.~\ref{fig:roc} shows that the profile-based curve dominates across most of the operating range and is particularly strong at low false-positive rates, the regime most relevant for deployment, where alert volume must remain manageable.

The ordering mirrors Experiment I: supplying raw agent context improves detection over the trajectory alone (+5.4 AUROC points), and structuring that context into an explicit Intent Profile provides a further gain of similar size (+5.2 points).

Because ground truth is derived from executed environment state rather than LLM judgment, and the benchmark was neither generated nor validated by our pipeline, this result provides independent evidence that the Intent Profile improves detection of behavioral drift, not only classification of requests.

\subsection{Experiment III: Intent Profile-based Input Alignment and Hardening}\label{sec:experiment-iii-intent-profile-based-input-alignment-and-hardening}

We evaluate the downstream behavioral effects of Input Alignment and System Prompt Hardening under a common simulation-and-evaluation protocol. For Input Alignment, we compare (1) an unprotected baseline that retains the original system prompt, request, memory, and tools; (2) DataFilter~\cite{ref16}; and (3) Intent Profile-based Input Alignment. Both alignment methods transform the user request, historical conversation, and tool responses.

For System Prompt Hardening, we compare the same baseline with (1) Polymorphic Prompt Assembling (PPA)~\cite{ref15}; (2) StruQ-style prompt-format hardening~\cite{ref10}; and (3) Intent Profile-based Hardening. The StruQ-style baseline adopts StruQ's structured-query prompt format to separate trusted system instructions from untrusted content, but does not implement StruQ's front-end filtering or structured instruction tuning. Each hardening method produces an effective system prompt.

The Intent Profile-based methods receive the Intent Profile. DataFilter uses the original system prompt as trusted instruction, whereas PPA and the StruQ-style prompt-format baseline receive no additional trusted context.

To evaluate downstream behavior consistently across conditions, we use a simulator to generate an agent response and a separate evaluator to assess whether the original request was achieved or refused. For each benchmark sample, the simulator receives the effective system prompt, current user request, conversation history, available tool responses, and declared tools, and produces a plan or agent answer. The response evaluator receives the original request, historical conversation, prior tool responses, and simulated response, but not the benchmark label. It determines whether the original request was fully achieved, partially refused, or fully refused.

For Out-of-Scope, Abusive, and Malicious samples, success requires full refusal; for Benign samples, success requires full achievement. Thus, any fulfillment of an unintended request is a failure. This shared protocol measures both security behavior and preservation of legitimate utility across all conditions.

We use MAI-Thinking-1 for profile extraction, hardening, and Intent Profile-based Input Alignment; GPT-5.6-luna~\cite{ref40} for simulation; and DeepSeek-V4-Pro for response evaluation. Against 100 manually reviewed samples, the response evaluator achieved 95\% accuracy.

\paragraph{Results}\label{sec:results-expIII}

Table~\ref{tab:exp3-alignment} reports success rate of each protection method. The success rate is based on the pass/fail classification produced by the evaluator.

When comparing Intent Profile-based Input Alignment with DataFilter we can see a greater overall advantage in aligning the data of 7 points. It is also much stronger in Out-of-Scope and Abusive classes. While Intent Profile-based Input Alignment achieved the best overall score, it had a decrease in points versus the baseline LLM on Benign class. Details in the Table~\ref{tab:exp3-alignment}.

\begin{table}[t]
\centering
\caption{Protection success rate (\%) of input-alignment methods.}
\label{tab:exp3-alignment}
\small
\setlength{\tabcolsep}{3pt}
\begin{tabular}{@{}lccccc@{}}
\toprule
 & \textbf{Macro} & \textbf{Benign} & \textbf{OoS} & \textbf{Abus.} & \textbf{Mal.} \\
\midrule
Baseline LLM & 69 & \textbf{97} & 49 & 85 & 44 \\
DataFilter & 78 & 90 & 67 & 90 & 66 \\
Profile-based Alignment & \textbf{85} & 91 & \textbf{85} & \textbf{96} & 66 \\
\bottomrule
\end{tabular}
\end{table}

Intent Profile-based Hardening achieves the highest macro protection success (83\%, vs. 81\% for PPA and 80\% for the StruQ-style prompt-format baseline), a small margin. Its clearest gain is on Out-of-Scope interactions (77\% vs. 72\% and 67\%). It is slightly higher on Abusive (96\% vs. 94\% and 91\%), and matches the StruQ-style prompt-format baseline on Malicious (64\%). It is slightly worse on Benign samples (94\% vs. 97\%), indicating a modest over-refusal cost. Full results are in Appendix~\ref{app:hardening}.

\subsection{Overall findings}\label{sec:overall-findings}

Our experiments provide three findings. First, structuring trusted agent context as an Intent Profile improves four-class macro-F1 by 34 points over raw context, including a 68-point gain in Out-of-Scope F1. Second, on executed AgentDojo trajectories with environment-derived ground truth, profile-based post-execution detection reaches 94.9\% AUROC versus 89.7\% with raw configuration, and 90.9\% for AlignmentCheck. Third, profile-based input alignment improves protection success by 7 points over DataFilter, with the largest gains on Out-of-Scope and Abusive interactions.

\section{Discussion}\label{sec:discussion}

The Intent Profile's main contribution appears to be representational: it explicitly represents the relationship between an agent's purpose, permitted workflows, tools, data boundaries, and prohibited behaviors. This may help explain why its largest gains occur in tasks whose labels depend strongly on deployment context, particularly Out-of-Scope classification and alignment.

\textbf{Deployment considerations.} Running a large LLM on every agent input or trajectory would be prohibitively expensive in production. Intent-Aware protection separates the work into two parts with very different costs. The expensive step, profile extraction, runs once per agent and only needs to be repeated when the agent's configuration or governing policies change, which is infrequent. The detection step runs on every interaction, but it receives the already-structured Intent Profile instead of raw configuration and policy documents. That makes it a narrower task, suited to a smaller, faster model. We hypothesize that this narrower task can be served by a smaller, lower-latency model; quantifying this accuracy--cost trade-off is future work. This split confines the expensive step to slowly changing agent configurations and could allow the high-volume stream of runtime events to be handled at lower cost.

\textbf{Limitations.} Our benchmark represents organizational and role intent as synthetic text summaries rather than production policy corpora, and its samples are realized and validated by LLMs. We mitigate this by fixing ground-truth labels before generation, using a different model for validation than for generation, and manually reviewing a random sample of 100 accepted items (98\% correctly labeled). Experiment III measures simulated rather than executed agent behavior and relies on an LLM evaluator (95\% agreement with manual review).

The AgentDojo evaluation provides executed, deterministically labeled trajectories, but covers only four agents, a single agent model (GPT-4o), and 97 benign runs, so most negatives are failed attacks rather than benign executions. Moreover, AgentDojo treats the user request as trusted and contains only indirect prompt injections, so it covers only the injection-based Malicious portion of our threat model. Out-of-Scope requests, Abusive repurposing, and user-side attacks are evaluated only on our synthetic, LLM-validated benchmark.

Finally, our experiments depend on closed-weights models, which may limit exact reproducibility.

\section{Conclusion}\label{sec:conclusion}

This work introduces Intent-Aware Security as a framework for protecting AI agents according to their trusted organizational, role, and developer intent. We formalize Intent Preservation, introduce the Intent Profile as an explicit representation of trusted agent intent, and operationalize that representation through four Intent Profile-based protection mechanisms: pre-execution detection, post-execution detection, system-prompt hardening, and input alignment. Together with the generative Intent-Aware benchmark, these mechanisms extend agent protection beyond attack signatures and permission checks toward security decisions conditioned on what a specific agent was deployed to do.

Our evaluation shows that explicitly structuring trusted agent context as an Intent Profile substantially improves fine-grained interaction classification and strengthens profile-conditioned protection, particularly for Out-of-Scope and Abusive interactions. These results suggest that intended purpose can serve as a practical security primitive for distinguishing legitimate use, non-adversarial scope violations, deliberate repurposing, and explicit attacks. The detection gains also hold for post-execution monitoring on an external benchmark with real tool execution and environment-derived ground truth.

\appendices
\section{Confusion-Difference Matrix for Experiment~I}\label{app:confusion}
Let \(B\) denote the no-profile confusion matrix and \(P\) the profile-based confusion matrix. Because support differs per class, each matrix is first normalized by ground-truth class:

\[\widehat{B}_{ij} = \frac{B_{ij}}{\sum_{j} B_{ij}} \times 100, \quad \widehat{P}_{ij} = \frac{P_{ij}}{\sum_{j} P_{ij}} \times 100
\]

The difference matrix is then:

\[C_{ij} = \widehat{P}_{ij} - \widehat{B}_{ij}
\]

Thus, each cell represents the percentage-point change in predictions after introducing the Intent Profile.

\begin{table}[h]
\centering
\caption{Confusion-difference matrix $C$ (percentage points), Profile-based vs.\ Raw-context classifier. Rows: ground truth; columns: predicted label.}
\label{tab:confusion-diff}
\small
\setlength{\tabcolsep}{4pt}
\begin{tabular}{@{}lcccc@{}}
\toprule
\textbf{GT $\downarrow$ / Pred.\ $\rightarrow$} & \textbf{Benign} & \textbf{OoS} & \textbf{Abus.} & \textbf{Mal.} \\
\midrule
\textbf{Benign} & \textbf{+7.2} & +8.4 & $-$16.0 & +0.4 \\
\textbf{Out-of-Scope} & \textbf{$-$56.1} & \textbf{+74.0} & $-$22.2 & +4.1 \\
\textbf{Abusive} & $-$0.4 & +11.0 & \textbf{$-$15.9} & +5.3 \\
\textbf{Malicious} & \textbf{$-$6.7} & +6.6 & \textbf{$-$41.5} & \textbf{+41.6} \\
\bottomrule
\end{tabular}
\end{table}

For diagonal cells, positive values indicate improved correct classification, while negative values indicate reduced correct classification. For off-diagonal cells, negative values generally indicate reduced confusion.

\section{Intent Profile-Based Hardening Results}\label{app:hardening}
Table~\ref{tab:hardening} reports the results of Intent Profile-Based Hardening.

\begin{table}[h]
\centering
\caption{Protection success rate (\%) of system-prompt hardening methods.}
\label{tab:hardening}
\small
\setlength{\tabcolsep}{3pt}
\begin{tabular}{@{}lccccc@{}}
\toprule
 & \textbf{Macro} & \textbf{Benign} & \textbf{OoS} & \textbf{Abus.} & \textbf{Mal.} \\
\midrule
Baseline LLM & 69 & 97 & 49 & 85 & 44 \\
PPA & 81 & 97 & 72 & 94 & 62 \\
StruQ-style format & 80 & 97 & 67 & 91 & 64 \\
Profile-based Hardening & \textbf{83} & 94 & \textbf{77} & \textbf{96} & 64 \\
\bottomrule
\end{tabular}
\end{table}


\begin{thebibliography}{99}

\bibitem{ref1}
K. Greshake, S. Abdelnabi, S. Mishra, C. Endres, T. Holz, and M. Fritz, ``Not What You've Signed Up For: Compromising Real-World LLM-Integrated Applications with Indirect Prompt Injection,'' in \emph{Proc. 16th ACM Workshop on Artificial Intelligence and Security (AISec)}, 2023, pp. 79--90, doi: 10.1145/3605764.3623985.

\bibitem{ref2}
Q. Zhan, Z. Liang, Z. Ying, and D. Kang, ``InjecAgent: Benchmarking Indirect Prompt Injections in Tool-Integrated Large Language Model Agents,'' in \emph{Findings of the Association for Computational Linguistics: ACL 2024}, Bangkok, Thailand: Association for Computational Linguistics, 2024, pp. 10471--10506, doi: 10.18653/v1/2024.findings-acl.624.

\bibitem{ref3}
H. Zhang, J. Huang, K. Mei, Y. Yao, Z. Wang, C. Zhan, H. Wang, and Y. Zhang, ``Agent Security Bench (ASB): Formalizing and Benchmarking Attacks and Defenses in LLM-Based Agents,'' in \emph{Proc. International Conference on Learning Representations (ICLR)}, 2025. [Online]. Available: \url{https://arxiv.org/abs/2410.02644}. Accessed: Nov. 6, 2026.

\bibitem{ref4}
Z. Deng, Y. Guo, C. Han, W. Ma, J. Xiong, S. Wen, and Y. Xiang, ``AI agents under threat: A survey of key security challenges and future pathways,'' arXiv:2406.02630, 2024, doi: 10.48550/arXiv.2406.02630.

\bibitem{ref5}
A. Chhabra, S. Datta, S. K. Nahin, and P. Mohapatra, ``Agentic AI security: Threats, defenses, evaluation, and open challenges,'' \emph{IEEE Access}, vol. 14, pp. 49455--49482, 2026, doi: 10.1109/ACCESS.2026.3675554.

\bibitem{ref6}
OWASP Foundation, \emph{OWASP Top 10 for LLM Applications 2026}, ver. 1.0, Aug. 2026. [Online]. Available: \url{https://genai.owasp.org/resource/owasp-genai-llm-top-10-2026/}. Accessed: Sep. 25, 2026.

\bibitem{ref7}
OWASP Foundation, \emph{OWASP Top 10 for Agentic Applications for 2026}, ver. 2026, Dec. 2025. [Online]. Available: \url{https://genai.owasp.org/resource/owasp-top-10-for-agentic-applications-for-2026/}. Accessed: Nov. 6, 2026.

\bibitem{ref8}
MITRE, ``MITRE ATLAS: Adversarial Threat Landscape for Artificial-Intelligence Systems.'' [Online]. Available: \url{https://atlas.mitre.org/}. Accessed: Nov. 6, 2026.

\bibitem{ref9}
K. Hines, G. Lopez, M. Hall, F. Zarfati, Y. Zunger, and E. K{\i}c{\i}man, ``Defending Against Indirect Prompt Injection Attacks with Spotlighting,'' arXiv:2403.14720, 2024.

\bibitem{ref10}
S. Chen, J. Piet, C. Sitawarin, and D. Wagner, ``StruQ: Defending Against Prompt Injection with Structured Queries,'' in \emph{Proc. 34th USENIX Security Symposium (USENIX Security 25)}, Seattle, WA, USA: USENIX Association, 2025, pp. 2383--2400. [Online]. Available: \url{https://www.usenix.org/system/files/usenixsecurity25-chen-sizhe.pdf}. Accessed: Nov. 6, 2026.

\bibitem{ref11}
E. Wallace, K. Xiao, R. Leike, L. Weng, J. Heidecke, and A. Beutel, ``The Instruction Hierarchy: Training LLMs to Prioritize Privileged Instructions,'' arXiv:2404.13208, 2024.

\bibitem{ref12}
S. Chen, A. Zharmagambetov, S. Mahloujifar, K. Chaudhuri, D. Wagner, and C. Guo, ``SecAlign: Defending Against Prompt Injection with Preference Optimization,'' in \emph{Proc. ACM SIGSAC Conference on Computer and Communications Security (CCS)}, 2025, doi: 10.1145/3719027.3744836.

\bibitem{ref13}
J. Yi, Y. Xie, B. Zhu, E. K{\i}c{\i}man, G. Sun, X. Xie, and F. Wu, ``Benchmarking and Defending Against Indirect Prompt Injection Attacks on Large Language Models,'' in \emph{Proc. 31st ACM SIGKDD Conference on Knowledge Discovery and Data Mining (KDD)}, 2025, doi: 10.1145/3690624.3709179.

\bibitem{ref14}
H. Li, X. Liu, N. Zhang, and C. Xiao, ``PIGuard: Prompt Injection Guardrail via Mitigating Overdefense for Free,'' in \emph{Proc. 63rd Annual Meeting of the Association for Computational Linguistics, Volume 1: Long Papers}, 2025, pp. 30420--30437, doi: 10.18653/v1/2025.acl-long.1468.

\bibitem{ref15}
Z. Wang, N. Nagaraja, L. Zhang, H. Bahsi, P. Patil, and P. Liu, ``To Protect the LLM Agent Against the Prompt Injection Attack with Polymorphic Prompt,'' in \emph{Proc. IEEE/IFIP International Conference on Dependable Systems and Networks Workshops (DSN-W)}, 2025, pp. 22--28, doi: 10.1109/DSN-S65789.2025.00037.

\bibitem{ref16}
Y. Wang, S. Chen, R. Alkhudair, B. Alomair, and D. Wagner, ``Defending Against Prompt Injection with DataFilter,'' in \emph{Proc. IEEE Conference on Secure and Trustworthy Machine Learning (SaTML)}, 2026, pp. 139--157. [Online]. Available: \url{https://arxiv.org/abs/2510.19207}. Accessed: Nov. 6, 2026.

\bibitem{ref17}
E. Debenedetti, I. Shumailov, T. Fan, J. Hayes, N. Carlini, D. Fabian, C. Kern, C. Shi, A. Terzis, and F. Tramèr, ``Defeating Prompt Injections by Design,'' in \emph{Proc. IEEE Conference on Secure and Trustworthy Machine Learning (SaTML)}, 2026. [Online]. Available: \url{https://arxiv.org/abs/2503.18813}. Accessed: Nov. 6, 2026.

\bibitem{ref18}
M. Costa, B. Köpf, A. Kolluri, A. Paverd, M. Russinovich, A. Salem, S. Tople, L. Wutschitz, and S. Zanella-Béguelin, ``Securing AI Agents with Information-Flow Control,'' arXiv:2505.23643, 2025.

\bibitem{ref19}
S. A. Siddiqui, R. Gaonkar, B. Köpf, D. Krueger, A. Paverd, A. Salem, S. Tople, L. Wutschitz, M. Xia, and S. Zanella-Béguelin, ``Permissive Information-Flow Analysis for Large Language Models,'' \emph{Transactions on Machine Learning Research}, 2025. [Online]. Available: \url{https://mlanthology.org/tmlr/2025/siddiqui2025tmlr-permissive/}. Accessed: Nov. 6, 2026.

\bibitem{ref20}
H. Wang, C. M. Poskitt, and J. Sun, ``AgentSpec: Customizable Runtime Enforcement for Safe and Reliable LLM Agents,'' in \emph{Proc. 48th IEEE/ACM International Conference on Software Engineering (ICSE)}, 2026, doi: 10.1145/3744916.3764546.

\bibitem{ref21}
K.-H. Hung, C.-Y. Ko, A. Rawat, I.-H. Chung, W. H. Hsu, and P.-Y. Chen, ``Attention Tracker: Detecting Prompt Injection Attacks in LLMs,'' arXiv:2411.00348, 2024.

\bibitem{ref22}
Y. Jiang, X. Gao, T. Peng, Y. Tan, X. Zhu, B. Zheng, and X. Yue, ``HiddenDetect: Detecting Jailbreak Attacks Against Large Vision-Language Models via Monitoring Hidden States,'' in \emph{Proc. 63rd Annual Meeting of the Association for Computational Linguistics, Volume 1: Long Papers}, 2025, pp. 14880--14893.

\bibitem{ref23}
S. Abdelnabi, A. Fay, G. Cherubin, A. Salem, M. Fritz, and A. Paverd, ``Get My Drift? Catching LLM Task Drift with Activation Deltas,'' in \emph{Proc. IEEE Conference on Secure and Trustworthy Machine Learning (SaTML)}, Copenhagen, Denmark, 2025, pp. 43--67. [Online]. Available: \url{https://arxiv.org/abs/2406.00799}. Accessed: Nov. 6, 2026.

\bibitem{ref43}
S. Chennabasappa, C. Nikolaidis, D. Song, D. Molnar, S. Ding, S. Wan, S. Whitman, \emph{et al.}, ``LlamaFirewall: An open source guardrail system for building secure AI agents,'' arXiv:2505.03574, May 2025.

\bibitem{ref44}
F. Jia, T. Wu, X. Qin, and A. Squicciarini, ``The Task Shield: Enforcing task alignment to defend against indirect prompt injection in LLM agents,'' in \emph{Proc. 63rd Annu. Meeting Assoc. Comput. Linguistics (ACL)}, Vienna, Austria, 2025, pp. 29680--29697, doi: 10.18653/v1/2025.acl-long.1435.

\bibitem{ref45}
K. Zhu, X. Yang, J. Wang, W. Guo, and W. Y. Wang, ``MELON: Provable defense against indirect prompt injection attacks in AI agents,'' in \emph{Proc. 42nd Int. Conf. Mach. Learn. (ICML)}, PMLR vol. 267, Vancouver, Canada, 2025. [Online]. Available: \url{https://proceedings.mlr.press/v267/zhu25z.html}. Accessed: Nov. 6, 2026.

\bibitem{ref24}
R. Arike, E. Donoway, H. Bartsch, and M. Hobbhahn, ``Evaluating Goal Drift in Language Model Agents,'' in \emph{Proc. Eighth AAAI/ACM Conference on AI, Ethics, and Society (AIES)}, vol. 8, no. 1, 2025, pp. 192--203, doi: 10.1609/aies.v8i1.36541.

\bibitem{ref25}
A. Menon, M. Saebo, T. Crosse, S. Gibson, E. Jang, and D. Cruz, ``Inherited Goal Drift: Contextual Pressure Can Undermine Agentic Goals,'' in \emph{ICLR 2026 Workshop on Lifelong Agents: Learning, Aligning, Evolving}, 2026; also available as arXiv:2603.03258.

\bibitem{ref26}
V. Siu, J. He, K. Montgomery, Z. Wang, N. Gong, C. Wang, and D. Song, ``A Framework for Formalizing LLM Agent Security,'' arXiv:2603.19469, 2026, doi: 10.48550/arXiv.2603.19469.

\bibitem{ref27}
L. Tsai and E. Bagdasarian, ``Contextual Agent Security: A Policy for Every Purpose,'' in \emph{Proc. Workshop on Hot Topics in Operating Systems (HotOS '25)}, Banff, AB, Canada, 2025, doi: 10.1145/3713082.3730378.

\bibitem{ref28}
E. Debenedetti, J. Zhang, M. Balunovi\'{c}, L. Beurer-Kellner, M. Fischer, and F. Tramèr, ``AgentDojo: A Dynamic Environment to Evaluate Prompt Injection Attacks and Defenses for LLM Agents,'' in \emph{Advances in Neural Information Processing Systems}, vol. 37, 2024. [Online]. Available: \url{https://proceedings.neurips.cc/paper_files/paper/2024/file/97091a5177d8dc64b1da8bf3e1f6fb54-Paper-Datasets_and_Benchmarks_Track.pdf}. Accessed: Nov. 6, 2026.

\bibitem{ref29}
I. Evtimov, A. Zharmagambetov, A. Grattafiori, C. Guo, and K. Chaudhuri, ``WASP: Benchmarking Web Agent Security Against Prompt Injection Attacks,'' in \emph{Advances in Neural Information Processing Systems}, 2025. [Online]. Available: \url{https://proceedings.nips.cc/paper_files/paper/2025/file/1c9818387f5dd0a0bc151214660f059d-Paper-Datasets_and_Benchmarks_Track.pdf}. Accessed: Nov. 6, 2026.

\bibitem{ref30}
S. Weng, Y. Feng, J. Zhang, X. Xie, J. Yu, and J. Liu, ``ARGUS: Defending LLM Agents Against Context-Aware Prompt Injection,'' arXiv:2605.03378, 2026, doi: 10.48550/arXiv.2605.03378.

\bibitem{ref31}
Z. Wang, Y. Li, Y. Wu, K. Chen, Z. Liu, F. K. Wai, P.-Y. Chen, V. L. L. Thing, B. Li, D. Tao, and T. Zhang, ``Who Pays the Price? Stakeholder-Centric Prompt Injection Benchmarking for Real-World Web Agents,'' arXiv:2606.13385, 2026, doi: 10.48550/arXiv.2606.13385.

\bibitem{ref32}
L. Ouyang \emph{et al}., ``Training Language Models to Follow Instructions with Human Feedback,'' in \emph{Advances in Neural Information Processing Systems}, vol. 35, 2022, pp. 27730--27744. [Online]. Available: \url{https://arxiv.org/abs/2203.02155}. Accessed: Nov. 6, 2026.

\bibitem{ref33}
Y. Bai \emph{et al}., ``Constitutional AI: Harmlessness from AI Feedback,'' arXiv:2212.08073, 2022.

\bibitem{ref34}
R. Greenblatt \emph{et al}., ``Alignment Faking in Large Language Models,'' arXiv:2412.14093, 2024.

\bibitem{ref35}
D. Dalrymple \emph{et al}., ``Towards Guaranteed Safe AI: A Framework for Ensuring Robust and Reliable AI Systems,'' \emph{arXiv preprint arXiv:2405.06624}, 2024.

\bibitem{ref36}
L. Derczynski, E. Galinkin, J. Martin, S. Majumdar, and N. Inie, ``garak: A framework for security probing large language models,'' \emph{arXiv preprint arXiv:2406.11036}, 2024.

\bibitem{ref37}
OpenAI, ``GPT-5.4 model,'' \emph{OpenAI API Documentation}, Mar. 2026. [Online]. Available: \url{https://developers.openai.com/api/docs/models/gpt-5.4}. Accessed: Sep. 25, 2026.

\bibitem{ref38}
Microsoft AI, ``MAI-Thinking-1: Building a Hill-Climbing Machine,'' technical report, Aug. 2026. [Online]. Available: \url{https://microsoft.ai/pdf/mai-thinking-1.pdf}. Accessed: Nov. 6, 2026.

\bibitem{ref39}
DeepSeek-AI, ``DeepSeek-V4: Towards highly efficient million-token context intelligence,'' arXiv:2606.19348, 2026. [Online]. Available: \url{https://huggingface.co/deepseek-ai/DeepSeek-V4-Pro}. Accessed: Sep. 25, 2026.

\bibitem{ref46}
OpenAI, ``GPT-4o model,'' \emph{OpenAI API Documentation}, May 2024. [Online]. Available: \url{https://developers.openai.com/api/docs/models/gpt-4o}. Accessed: Oct. 23, 2026.

\bibitem{ref40}
OpenAI, ``GPT-5.6 Luna model,'' \emph{OpenAI API Documentation}, Jul. 2026. [Online]. Available: \url{https://developers.openai.com/api/docs/models/gpt-5.6-luna}. Accessed: Sep. 25, 2026.

\end{thebibliography}
\end{document}